\documentclass[mathleft]{an}
\usepackage{graphicx}
\usepackage{times}

\newcommand\nat{\textit {Nature}}
\newcommand\aap{\textit {A\&A}}
\def\pref#1{(\ref{#1})}

\def\pref#1{(\ref{#1})}

\def\bar{\overline}
\def\OV{\overline{\bf V}}
\def\OB{\overline{\bf B}}
\def\emf{\overline{\mbox{${\cal E}$}} {}}
\def\emfb{\overline{\mbox{\boldmath ${\cal E}$}} {}}
\def\bbE{\bar {\bf E}}

\def\beq{\begin{equation}}
\def\eeq{\end{equation}}
\def\lsim{\mathrel{\rlap{\lower4pt\hbox{\hskip1pt$\sim$}}
    \raise1pt\hbox{$<$}}}
\def\gsim{\mathrel{\rlap{\lower4pt\hbox{\hskip1pt$\sim$}}
    \raise1pt\hbox{$>$}}}
\def\bfE{{\bf E}}
\def\bfJ{{\bf J}}

\def\bfA{{\bf A}}
\def\bfa{{\bf a}}

\def\bfB{{\bf B}}
\def\bbJ{\bar {\bf J}}

\def\bB{\overline B}
\def\ts{\times}
\def\lb{\langle}
\def\rb{\rangle}
\def\curl{\nabla {\ts}}
\def\nt{\nabla\times}

\def\bfv{{\bf v}}

\def\bfj{{\bf j}}

\def\bfb{{\bf b}}

\def\bfB{{\bf B}}

\def\bbB{\overline {\bf B}}
\def\bbA{\overline {\bf A}}
\def\nt{\nabla\times}
\def\div{\nabla\cdot}
\def\OB{\overline{\bf B}}

\def\ob{\overline{B}}

\begin{document}

\Pagespan{789}{}
\Yearpublication{2010}%
\Yearsubmission{2009}%
\Month{11}%
\Volume{999}%
\Issue{88}%

\title{Comparisons and Connections between Mean Field Dynamo Theory and Accretion Disc Theory}

\author{E.G. Blackman \inst{}\fnmsep\thanks{Corresponding author:
  \email{blackman@pas.rochester.edu}\newline}
}
\titlerunning{Instructions for authors}
\authorrunning{E.G. Blackman }
\institute{
Department of Physics and Astronomy, University of Rochester, 
Rochester, NY, 14627, USA}

\received{30 May 2005}
\accepted{11 Nov 2005}
\publonline{later}


\abstract{The origin of large scale magnetic fields in astrophysical
rotators, and the conversion of gravitational energy into radiation near stars and compact objects via accretion have been  subjects of active research for a half century. Magnetohydrodynamic turbulence  makes both problems highly nonlinear, so both subjects have  benefitted from numerical simulations.However, understanding the key principles and  practical modeling of observations warrants testable  semi-analytic mean field theories that distill the essential physics.  Mean field dynamo (MFD) theory  and alpha-viscosity accretion disc theory
exemplify this pursuit.  That the latter is a mean field theory is not always made explicit but the combination of  turbulence and global symmetry imply such.    The more commonly explicit presentation of assumptions in  20th century textbook MFDT  has exposed it to arguably more widespread criticism than incurred by 20th century alpha-accretion theory despite complementary weaknesses.   In the  21st century however,  MFDT has experienced a breakthrough with a dynamical saturation theory  that consistently agrees with simulations. Such has not yet occurred in accretion disc theory, though progress is emerging. Ironically however, for accretion engines,  MFDT and accretion theory are presently two artificially uncoupled pieces of what should be a single coupled theory. Large scale fields and accretion flows are dynamically intertwined 
because large scale fields likely play a key role in angular momentum transport. I  
discuss and synthesize aspects of recent  progress in MFDT  and  accretion disc theory 
to  suggest why the two likely conspire in a unified theory.}
\maketitle

\section{Introduction}

Large scale dynamo (LSD) theory  in astrophysics is aimed at quantitatively understanding the in situ physics of magnetic field  growth, saturation, and  sustenance  on time or spatial scales large compared to the turbulent scales of the host.
 The presence of magnetohydrodynamic  turbulence makes the problem highly nonlinear, exacerbating  the importance of numerical simulations.  However, large scale spatial symmetry and slow evolution of  large scale fields  compared to turbulent fluctuations  time scales motivates  a mean field approach in which statistical, spatial, or temporal averages are taken  and evolution of the mean field studied (Moffatt 1978). An important goal of 21st century semi-analytic mean field dynamo theory  (MFDT) is to capture the essential  nonlinear physics in a minimalist way that is simple enough for phenomenological models, but also accounts for  the nonlinear saturation found in simulations  (for technical reviews see: Brandenburg \& Subramanian 2005; Blackman 2007).  The past decade has revealed significant progress in this endeavor as the  growth and saturation of LSDs seen in simulations seems to be reasonably matched by mean field dynamo (MFD) theories that follow the time dependent dynamical evolution of magnetic helicity (e.g. Blackman \& Field 2002).

Stars show that in situ LSDs must operate in nature.  The sun exhibits sign reversals  of the large scale field over each half  solar cycle (e.g. Wang \& Sheely 1993;  Schrijver  \&  Zwaan 2000).   If the field were simply that frozen into to the ISM during initial formation, there would  be no reversals. LSDs are also likely  operating in galaxies, despite whatever initial fields may have been seeded cosmologically because  supernova driven turbulence requires sustenance of the large field against 
turbulent diffusion.  Turbulent diffusion in 3-D is not likely suppressed as it may be in 
in 2-D  simulations  for weak magnetic fields (Cattaneo 1994) and there is little evidence that turbulent diffusion in interstellar media is quenched.  Detailed modeling of the geometry of large scale fields can be matched by mean field theories (\cite{beck96,vallee04,shukurov02}), though incorporation of the nonlinear LSD principles of the past decade for idealized dynamos into more geometrically realistic dynamos of astrophysical rotators with realistic boundary conditions setting is an ongoing avenue of research (e.g. Sur et al. 2007,2009).  

Given the role of LSDs in the turbulent rotators of stars and galaxies, their presence  in turbulent accretion discs is also likely. 
The large scale fields required by prevailing models of astrophysical jets may be produced by an in situ dynamo, although understanding the relative importance of in situ production vs. flux accretion  in that context has been a long-standing topic of research (e.g. Lovelace et al. 2009). As in  stellar and galactic environments of LSDs, the likely  presence of  magnetohydrodynamic turbulence in accretion discs makes the problem  fundamentally nonlinear and a focus of numerical simulations.  
 Computational resources have demanded local shearing boxes for thin disc studies
 but, as discussed in section 4, even localized stratified shearing box simulations show evidence for LSDs and the influence of large scale fields (Brandenburg et al. 1995; Lesur \& Ogilive 2008; Davis et al. 2009).   Non-thermal coronal luminosity from accretion engines 
  also hints at the role of large scale field  because larger scale structures more easily survive the buoyant rise before dissipating in  coronae (Blackman \& Pessah 2009).

While LSD theory has always been presented with  explicit mean field theories that expose the approximations made, axisymmetric accretion disc models such as the Shakura Sunyaev  (\cite{ss73}, hereafter SS73) viscous $\alpha_{ss}$ model are not always recognized as mean field theories by those using them to model spectra.  In this sense, 21st century LSD theory is more progressed than axisymmetric mean field accretion disc theory:
The $\alpha_{ss}$ disc prescription  does not yet offer predictive power for numerical simulations, whereas 21st century LSD theory  predicts the saturation of LSDs seen in a range of  simulations.  Ironically, because large scale fields likely play a fundamental role in accretion discs, both MFD and accretion  theory are really two faces of a single coupled theory.

Herein, I  first explore some conceptual progress and underlying principles of 21st century LSD/MFD theory in broad brush strokes.
I then discuss parallels between large scale dynamo theory and accretion disc theory as separate
endeavors. Finally,  I bring the two together and identify evolving connections and the importance of integrating the two theories into one.

 \section{Overview of 21st Century LSD Theory}

For $\sim$ 50 years, 20th century textbook MFD theory (e.g. \cite {moffatt})
has lacked a  saturation theory to predict how strong the large scale fields get
before  quenching  via the back-reaction of the field on the driving flow. 
But substantial progress toward a nonlinear  mean field theory has emerged  in the 21st century  via a symbiosis between analytical and numerical work.  Coupling the dynamical evolution of magnetic helicity into the dynamo equations turns out to provide fundamental insight to the saturation seen in simulations. Whether this approach is the only way to understand the saturation is uncertain but it has been successful.

The connection between magnetic helicity and large scale dynamos is implicit in the classic EDQNM spectral model of helical MHD turbulence in Pouquet et al. (1976). 
In Kleeorin \& Ruzmaikin (1982)  an equation that,  in hindsight, couples the time evolution of mean magnetic helicity to the dynamo is present though not actually solved for the time evolution.
 The success of the coupling becomes dramatically evident only when the time evolution is studied. That exploration has led to the paradigm shift in MFD theory.   For a sampling of such papers  see  Blackman \& Field (2002);  Brandenburg \& Subramanian (2005);  Blackman (2007); K\"apyl\"a et al. 2008). 


\subsection{Globally Helical }

Most  work in LSD/MFD theory has focused on systems that are initially globally reflection asymmetric (GRA).   In such systems,  pseudoscalars, such as the hemispherically averaged product of angular velocity and density gradient imposed by the initial setup
facilitate e.g. a kinetic helicity  pseudoscalar, common to  standard textbooks (\cite{moffatt,parker})
``$\alpha_{dyn}$ effect'' of mean field dynamos.   However,  21st century MFD theory has revealed that it is the magnetic helicity which should be considered the  unifying quantity for all MFDs. 
 The initial GRA conditions and the potential pseudoscalars that arise are important
 primarily because they provide one way to sustain  
an electromotive force aligned with the mean magnetic field. This alignment is a source
of magnetic helicity or magnetic helicity flux.

 There are two  classes of GRA LSDs.
The first is flow driven helical dynamos (FDHD) which apply
 inside of astrophysical rotators. Here the initial energy is dominated by velocity flows and the field responds.  
These are potentially linked to coronae where a second type of LSD, the magnetically driven helical dynamo (MDHD) can operate.  A MDHD can also be 
described as dynamical magnetic relaxation (Blackman \& Field 2004) 
toward the "Taylor" state (Taylor 1986).
Note that the Taylor state is the lowest energy state to which a magnetically dominated configuration evolves. 
For a closed system,  the state is characterized by the magnetic helicity migrating to the largest scale available subject to boundary conditions.  In an accretion disc, the  MDHD would  characterize coronal relaxation of magnetic structures (fed by helical fields from below)
into larger (even jet mediating) scales in a magnetically dominated environment.
In this respect, the MDHD in disc-corona or star-corona  interfaces are  analogous to laboratory plasma dynamos (\cite{bellan,jiprager})
 that occur in reverse field pinches (RFPs) and Spheromaks with the corona
being the internal volume of these configurations and the disc or stellar surface being the boundary.

LSDs always involve some helical growth of the large scale field which is coupled
to a helical growth of smaller  scale fields of opposite sign
and/or a compensating helicity flux when boundary conditions allow.
When small scale magnetic or current helicity evolution is coupled to the large scale 
field growth, the simplest 21st century dynamo reveals that the mean field `dynamo $\alpha_{dyn}$ becomes the difference between kinetic helicity and 
current helicity:  For an $\alpha^2_{dyn}$ type FDHD simulated in a closed box (Brandenburg 2001), 
the current helicity builds up as the large scale field 
grows and quenches the FDHD, in accordance with MFD predictions (Blackman \& Field 2002). 

The effect of boundary or flux terms can vary depending on the sign and 
relative flux of small and large scale magnetic helicities. For a closed system   the buildup of small scale magnetic (and current) helicity 
is necessarily accompanied by small scale magnetic helicity buildup with the opposite sign,
which quenches the LSD to at most a rate determined by the dreadfully slow resistive dissipation of small-scale helicity.
If this "catastrophic" quenching occurs before enough large scale field is grown in the fast growth regime, 
 a preferential flux of  helicity of the opposite sign through a realistic astrophysical boundary
is desirable to  alleviate this quenching  and sustain further fast fast growth.  For GRA systems,
in each hemisphere, the large scale helicity builds with one sign and the small scale with the opposite sign so the flux of the small scale helicity is desired.(Blackman \& Field 2000; Vishniac Cho 2001; Sur et al. 2007; K\"apyl\"a et al 2008).
It is possible that even if the field grows to acceptable magnitudes without boundary fluxes,
the cycle period becomes resistively limited.  In this case the  helicity fluxes can unclog the cycle.
  
The actual mechanism by which small scale helicity might be preferentially ejected 
(rather  than larger scale magnetic helicity) depends on the astrophysics of  particular system. The physics of helicity ejection should emerge as part of the natural buoyant loss of magnetic flux from a rotator, and accordingly in simulations.
Numerical simulations  support these basic principles: when shear is present, and when  surfaces of constant shear align toward open boundaries 
 a specific non-vanishing magnetic helicity flux is allowed which alleviates the quenching
(Brandenburg \& Sandin 2004; K\"apyl\"a et al. 2008).

 For an MDHD, the system is first dominated instead by the 
current helicity and a growing kinetic helicity can act as the back-reaction 
(Blackman \& Field 2004). 
Both simple FDHDs and MDHDs are accessible within the same formalism, all unified by tracking magnetic helicity evolution, and aided by thinking of the field as ribbons rather than lines
(Blackman \& Brandenburg 2003).
More work on how the fields evolve from within
the rotator to produce the global scale fields in  coronae 
 is needed. Interestingly, although most observations probe coronare, most work on astrophysical dynamo theory has focused on the FDHD and not the MDHD.

\subsection{Not Globally Helical}


LSD action has also been observed in non GRA simulations 
(e.g. Yousef et al 08; Lesur \& Ogilvie 08; Brandenburg et al. 2008). 
The minimum global ingredients for this class of LSD seem to be 
 shear  plus turbulence  (Yousef et al. 2008).
The non GRA LSDs grow large scale fields on scales larger than the outer scale of turbulence but smaller than the global scale. Although there is no GRA,  in regions where the large scale field is coherent regions, there is a field aligned electromotive force (EMF), and thus  an intermediate scale source of magnetic helicity that may switch signs between coherence regions and globally averages to zero.  It may be that the non-GRA LSD action always involves a local helicity flux  between coherence regions.  It is therefore important to recognize that the absence of GRA
need NOT imply the irrelevance of magnetic helicity and its flux.  

In the absence of invoking GRA, a number of approaches that appeal to the anisotropy of
the turbulence due to shear have been studied, and can provide finite contributions to the EMF
on intermediate scales (e.g. Vishniac \& Brandenburg 1997 [stochastic $\alpha_{dyn}$ effect]; Blackman 1998a [turbulent small scale dynamo +  large scale shear] 
shear); Kleeorin \& Rogachevskii 2003 [shear current effect, though see Brandenburg et al 2008 and particularly Sridhar and Subramanian 2009ab which challenge its existence]  Schekochihin et al 2008 [turbulence + shear]).
The role of  some kind of helicity flux (e.g. Vishniac \& Cho 2001) between
 quasi-local sectors rather than globally is likely important, 
but the dominant mechanism and the consequences of the type of mean fields produced
by non-GRA LSDs remain  topics of desired research.



\section{Why Magnetic Helicity  Is Relevant}

Magnetic helicity has emerged as a key quantity in understanding LSD saturation but why should this be? Several concepts conspire to provide insight into why. 

\subsection{Standard textbook mean field dynamos necessarily generate
large scale magnetic helicity}
The
 mean field  $\OB$ satisfies the induction 
equation (Moffatt 1978; \cite{kr80}) 
\beq
{\partial\OB\over \partial t} = -c\curl\bbE,
\label{2.4a} 
\eeq
where
\beq
c\bbE=-\OV\times \OB  - \lb\bfv\ts\bfb\rb+
{\nu_M }\nt\OB,
\label{2.4aaa} 
\eeq
where the magnetic diffusivity $\nu_M\equiv {\eta c^2\over 4\pi}$  and 
$\eta$ is the resistivity.
The  turbulent electromotive force $\emfb\equiv \lb \bfv\times \bfb \rb $  can in general be expanded in terms of powers of spatial and temporal derivates of the mean magnetic field and mean velocity (or mean shear).  The lowest order contributions to $\emf$ from the mean magnetic field terms lead
to exponential growth and diffusion in 20th century textbooks   (and are the only contributions in the absence of a mean
velocity) are given by
\beq
\emf_i = \alpha_{ij}{\bB}_j-\beta_{ijk}\partial_j{\bB}_k.
\label{2.5aa}
\eeq

Using Maxwell's equations, the definitions  ${\curl \bfA }\equiv \bfB$, and $\bfJ\equiv {c\over 4\pi} \curl \bfB$, along with vector identities, the equation for the evolution of magnetic helicity density $H^M\equiv \lb \bfA\cdot\bfB \rb $ is  (see Bellan 2000; Blackman 2007)
\beq
\partial_t( \bfA\cdot\bfB) 
=-2\nu_M(\bfJ\cdot\bfB)
-\div( 2\Phi \bfB + \bfA \ts \partial_t\bfA),
\label{4a}
\eeq
where $\Phi$ satisfies $\bfE = \nabla \Phi - {1\over c}{d\bfA\over dt}$.
Following the analogous procedure for the mean and fluctuating
quantities, and using (\ref{2.4aaa}) and ((\ref{2.5aa}) gives for the large and small
scale quantities $H_1^M\equiv \lb\bbA\cdot\bbB\rb$ and $H_2^M\equiv \lb \bfa\cdot \bfb \rb$ respectively
\beq
\partial_t(\bbA\cdot\bbB)
=2\emfb\cdot\bbB
-2\nu_M \bbJ\cdot\bbB
-\div( 2{\overline \Phi}\ \bbB + \bbA \ts \partial_t\bbA)
\label{5a}
\eeq
and 
\beq
\partial_t\overline{ \bfa\cdot\bfb}
=-2\emfb\cdot\bbB-
2\nu_M\overline{\bfj\cdot\bfb}
-\div(\overline{2{\phi} \bfb} + \overline{\bfa\ts \partial_t\bfa}).
\label{6a}
\eeq

To highlight that standard large scale dynamos necessarily involve generation of  mean magnetic helicity, note that the minimalist versions of such dynamos invoke
$\alpha_{ij}= \alpha_{dyn}\delta_{ij}$ and $\beta_{ijk}=\nu_{M,T}\epsilon_{ijk}$ where $\alpha_{dyn}$ 
is a  pseudoscalar  and $\nu_{M,T}$ acts as a mean field scalar diffusion coefficient. (Note that in  anisotropic systems, $\beta_{ijk}$ may have negative components that might act to amplify
rather than diffuse--an ingredient in some of the models of  non GRA LSDs discussed in section 2.2.) The growth of the large scale field then involves
a finite $\emfb\cdot\OB$, which provides an equal and opposite
source term to the large scale and small scale magnetic helicity as seen from
(\ref{5a}) and (\ref{6a}). 
 In solving (\ref{2.4a}), standard 20th century textbook theory
focuses on an equation for the large scale field from which one can write
an equation like (\ref{5a}). But 20th century dynamo theory does not couple in
the evolution of the small scale magnetic helicity contained
in (\ref{6a}). 

In the simplest of 21st century dynamos (e.g. Blackman \& Field 2002),  
$\alpha_{dyn}\sim -\tau (\lb \bfv\cdot\curl \bfv \rb -\lb\bfb\cdot\curl\bfb\rb)\equiv  \alpha_{kin} + \alpha_{mag}$, where $\tau$ is of order a correlation time of dominant fluctuations. In  20th century theory only $\alpha_{kin}$ contributes to $\alpha_{dyn}$.  The contribution $\alpha_{mag}$ is proportional to the small scale  current helicity, which in turn is proportional to $H_2^M$ in the Coulomb gauge for a closed system. Therefore the time evolution of $\alpha_{dyn}$ is directly coupled to the time evolution of magnetic helicity. If  $H_1^M$  grows initially from $\alpha_{kin}$, then  $H_2^M$ grows of opposite sign which in turn quenches $\alpha_{dyn}$. As alluded to in our discussion of MDHDs above, it is also possible to drive the dynamo instead with $\alpha_{mag}$, such that $\alpha_{kin}$ is the quencher.
I come back to this point in section 5.  


A subtle aspect of the helicity density equations (\ref{5a}) and (\ref{6a}) is the 
issue of gauge invariance. In the absence of boundary flux terms, all terms are manifestly
gauge invariant.  In the present of flux terms, the gauge invariance is broken (e.g. Berger \& Field 1986).  Subramanian \& Brandenburg (2006) carefully construct a generalized local helicity density that reduces to the above equations in the absence of flux terms 
and has the same form in the presence of flux terms but with a redefinition of $H_1^M$,  $H_2^M$ and the flux terms that avoids the use of the vector potential. The lowest order terms in their
definition are similar to what is obtained from the Coulomb gauge but the corrections make
each individual  terms  in the equation gauge invariant with a local physical meaning.
That important construction puts the physical discussion of helicity density evolution  equations here and elsewhere on  firm ground, particularly if one seeks to measure a local  flux density as a physical quantity.  However, if one uses the helicity evolution equations as intermediary equations in a particular gauge and  can convert back to the magnetic field, the resulting magnetic field will always be gauge independent.

\subsection{Evolution of Total Magnetic Helicity vs. Total Magnetic Energy }

The Taylor state (Taylor 1986), mentioned in section 2.1, 
is typically derived by starting with a bounded, static magnetically dominated configuration and minimizing the magnetic energy subject to the  constraint that magnetic helicity is  conserved. The result is 
a configuration in which  the magnetic helicity evolves to the largest scale available subject to boundary conditions. This is NOT a dynamical calculation, but  a calculation of an end state.
In fact, the evolution to this end state is a dynamo--albeit a MDHD.
Recall that the MDHD is an LSD for which magnetic helicity is injected at small scales and (non-locally) inverse cascades to large scales (Bellan 2000; Blackman \& Field 2004). 

The concept of the Taylor state  hints at  a fundamental feature of magnetic helicity and its connection to large scale dynamos. Even a FDHD in a closed volume involves an inverse transfer of magnetic helicity to large scales (Pouquet et al. 1976;  Brandenburg 2001)
For a system with initially zero net magnetic helicity the FDHD segregates one sign to large scales (e.g. the box scale) and the other sign
to the  largest scale available to it (the forcing scale).
The system then achieves a  "doubly relaxed state" (Blackman 2003).

Key in deriving Taylor states is the assumption  that magnetic helicity
evolves slowly compared to magnetic energy.  This also emerges as  justifiable 
a posteriori for  numerical simulations of the FDHD  (e.g. Brandenburg 2001). It is therefore instructive to evaluate  how widely applicable this assumption can be expected to hold.
More specifically, one can ask for what range of magnetic energy and magnetic helicity spectra does the the magnetic helicity decay more slowly than magnetic energy?

Following Blackman (2004) and working in the Coulomb gauge,  we write the total magnetic energy density
\beq
\langle\bfB^2 \rangle=M=
\int_{k_0}^{k_{\nu_M}} M_k dk 
\eeq
where $k_0$ and $k_{\nu_M}$ are the minimum and maximum (resistive) 
wave numbers, and where the magnetic energy density spectrum
\beq
M_k\equiv \int |{\tilde {\bf B}}|^2k^2d\Omega_k  =
\int |{\tilde {\bf A}}|^2k^4d\Omega_k 
\propto k^{-q}.
\eeq
Here $\Omega_k$ is the solid angle in wave-number space,   $q$ is assumed constant, 
and the tilde indicate Fourier transforms.
The magnetic helicity density
spectrum is  then
\beq
\begin{array}{r}
H_k\equiv {1\over 2}\int\left[{\tilde {\bf A}}(k) {\tilde {\bf B}}^*(k) + {\tilde {\bf A}}^*(k) {\tilde {\bf B}}(k)\right]k^2d\Omega_k\\ 
= M_k f(k)/k,
\end{array}
\eeq
where $f(k)\propto k^{-s}$ is the fraction of magnetic energy that is helical at each wave number
and $s$ is taken as a constant.
We then also have correspondingly
\beq
\langle\bfA\cdot \bfB \rangle=\int_{k_0}^{k_{\nu_M}}f(k)M_k k^{-1}dk
\label{10}
\eeq
and the current helicity
\beq
\langle\bfJ\cdot \bfB \rangle=\int_{k_0}^{k_{\nu_M}}f(k)k M_k dk.
\label{11}
\eeq

Now, using (\ref{5a}) for a closed system we have
\beq
\partial_t H=\partial_t\langle\bfA\cdot\bfB\rangle =-2{\nu_M}
 \lb \bfJ\cdot \bfB\rb
\label{12}
\eeq
and the magnetic induction equation gives
\beq
\partial_t M=\partial_t\langle\bfB^2\rangle =-2{\nu_M}\langle(\nabla \bfB)^2\rangle
-2 \langle {\bf v }\cdot (\bfJ\times \bfB)\rangle.
\label{13}
\eeq
Using (\ref{10}), (\ref{11}), (\ref{12}) and 
(\ref{13}) we then obtain
\beq
\tau_H= {-H\over \partial_t H}  =  {\int_{k_L}^{k_{\nu_M}} f(k) M_k k^{-1} dk \over 
2{\nu_M} \int_{k_L}^{k_{\nu_M}}f(k)kM_kdk}
\eeq
and
\beq
\tau_M= {-M\over (\partial_t M)_{res}} =  {\int_{k_L}^{k_{\nu_M}} M_k  dk \over 
2{\nu_M} \int_{k_L}^{k_{\nu_M}}k^2M_kdk}, 
\eeq
where the subscript "{\it res}" indicates the contribution from the penultimate term in (\ref{13})
only.  The range of  $s$ and $q$ for which 
 $R\equiv {\tau_H\over \tau_M}  > 1$ corresponds to regime in which 
 the magnetic helicity decays more slowly than the magnetic energy.
Blackman (2004) showed that $R>1$ for the combination of  $s>0$ and $3> q>0$.
and that $R<1$  for small $0 < q< 1$ and  $s <0 $.
Standard derivations  of the Taylor state are correct only when 
 $R>1$.  Indeed   $R>1$  for a wide range of reasonable spectra
and magnetic helicity is therefore often conserved better than magnetic energy.
 
Two key implications of this last statement are that: (1) the evolution of magnetic helicity in a system, its separation of scales, its flow out of a boundary, and its rate of dissipation
should be tracked in dynamo models and is missing from 20th century textbook dynamos which
do not conserve helicity (see  Fig 1. of Blackman \& Brandenburg  2003); and 
(2) Magnetic helicity conservation should really be considered 
 a more fundamental  constraint on closed box numerical MHD simulations than 
 magnetic energy conservation. This also applies to shearing box simulations (section 4).

 

\section{Accretion Theory: Role of Mean Field Theory and Large Scale Fields}

\subsection{Standard accretion theory is  a mean field theory}

Gaseous accretion discs around astrophysical rotators have long been studied as
a source of luminosity for compact engines (see Treves et al. 1989; Frank et al. 2002).
In astronomy,  a  practical 
semi-analytic formalism to be incorporated into emission and spectral models for comparison with observations is a desired goal.   Because of this goal, 
the presentation of practical accretion models
in standard texts  (e.g. Frank et al. 2002)   
 based on SS73  employs the 
Navier-Stokes equation and replaces the microphysical viscosity with a turbulent viscosity
without explaining the theoretical complexity and assumptions underpinning 
this  bold procedure.    The replacement of the microphysical viscosity by a turbulent viscosity
is itself a turbulent closure and warrants a formal derivation analogous to that 
of the MFD theory coefficients which
also require closures.  Closures approximate an otherwise infinite set of nonlinear turbulence equations by an approximation that facilitates a finite set.

That the SS73  model is both axisymmetric AND involves a turbulent viscosity implies that the theory is a mean field theory. Local axisymmetry never applies
on the scale of turbulent eddies.  The sense in which accretion disc theory is a mean
field theory is addressed in a subset of theoretically oriented literature
 (e.g. Balbus et al. 1994; Blackman 1997; Balbus \& Hawley 1998; 
 Ogilvie 2003 Pessah et al. 2006; Hubbard \& Blackman 2009). 
 In contrast,  20th century textbook LSD theory is  typically more explicitly presented  as a  mean field theory (e.g. Moffatt 1978, Parker 1979). This has arguably led to the 20th MFD theory being subject to broader critical scrutiny than SS73 despite similar weaknesses. 
MFD theory has traditionally focused on the magnetic induction equation as the "primary equation" of the theory and accretion disc models focus on the mass conservation and angular momentum transport equations as the "primary equations" of the theory with the coupling to the magnetic field
often ignored or "swept" into "$\alpha_{ss}$".

A practical   consequence of accretion theory being a mean field theory is 
a built in limitation in predictive precision associated with the stochasticity 
(Blackman 1998b).   For observations that collect data on time scales shorter than the largest eddy turnover time divided by the square root of the number of eddy cells contained in a spatially unresolved region contributing emission within a given frequency range, a mean field  theory  does not  make an exact prediction.  There is an intrinsic variability associated with predictions from any mean field theory. One has to be sure that the interpretation and use the theory corresponds with its correct meaning when compared to a given set of observations.

At present, the gap between formal identification of the minimal set of 
transport coefficients that best captures the angular momentum transport
from turbulence in accretion theory seen in simulation 
 and practical models for use in spectral modeling, 
  is larger than the analogous gap in MFD theory.
MFD theory now has a predictive semi-analytic theory that agrees with 
a set of nonlinear simulations which is not the case for mean field accretion theory like
SS73.    MFD theory has presently 2 tensor  coefficients
in common use (the $\alpha_{ij}$ and $\beta_{ijk}$ coefficients
which are  used even in semi-analytic dynamo models to match observed field structures in stars or in galaxies. Dynamical equations for these quantities that incorporate magnetic helicity evolution and the "minimal $\tau$" closure (e.g. Blackman \& Field 2002, see also Snellman et al. 2009 applied to shear flows) capture  MFD saturation. 
In accretion theory, there are few models that use anything other than  the single scalar diffusion coefficient of the SS73 formalism.  The minimum properties that 
turbulence must have to transport angular momentum outward are more subtle than what purely isotropic turbulence provides, and it is likely that more than one transport coefficient  is needed (Pessah et al. 2006; Hubbard \& Blackman 2009). Progress toward capturing the saturated stress of  the magneto-rotational instability (MRI) as a mean field model using a closure similar to the "minimal $\tau$"  approximation and going beyond
the $\alpha_{ss}$ disc model is emerging (Ogilvie 2003; Pessah et al. 2006).

Although the  magneto-rotational instability (MRI) has emerged 
as a leading candidate for local angular momentum transport in  accretion discs (e.g. Balbus \& Hawley 1998; Balbus 2003),  there remains  a disconnect between what 
shearing box simulations have told us about the 
MRI vs. how the instability might operate in nature.  To date, simulations have told us  that the MRI plausibly operates in accretion disc but not robust scalings of transport coefficients that spectral modelers can use. Patience (likely for for several more decades) is required, 
as the computational and conceptual demands are substantial.

To achieve maximal resolution, 
most first generation MRI simulations (except Brandenburg et al.1995)
did not use explicit viscosity or magnetic diffusivity.
Indeed recent work does show that for unstratified simulations
there is significant magnetic Prandtl number dependence
(Fromang et al. 2007), though this dependence is reduced for stratified boxes (Davis et al. 2009; Shi et al. 2009).
Similarly, the  angular momentum transport coefficient $\alpha_{ss}$ depends strongly on the box size and the strength of the initially imposed weak mean field strength for unstratified 
boxes with imposed mean fields (Pessah et al. 2007). 
Interestingly, $\alpha_{ss}$ varies $\sim 4$ orders between simulations, 
but $\alpha_{ss}\beta_p$, where $\beta_p$ is the ratio of thermal to magnetic pressure, 
is nearly a constant (Blackman et al. 2007).  
A theory that explains how the MRI
saturation depends upon all these choices is lacking, thereby limiting our quantitative implications
for understanding angular momentum transport in a realistic system. 
For unstratified boxes,  the 
dependence of the MRI saturation on parasitic modes and the need to have
large enough boxes to accommodate them is possibly quite  important (Pessah \& Goodman 2009). The mean field approach  of  Pessah et al. (2006)  hints at what minimal set of coefficients might be required in a more robust but still practical model of MRI mediated accretion discs.  
The approach better matches  the stress dependence on rotation profile than the $\alpha_{ss}$ of SS73   in numerical simulations of unstratified boxes with an initial mean field
 (Pessah et al. 2008).
.


\subsection{Why the focus on local transport? }

Accretion disc theory highlights is a duality of 
influence of computational tools.   

The practicality of the SS73 model
has focused attention toward local models of angular momentum transport.
Indeed, for much of the past 15 years,  the promising MRI has been studied as a local instability 
or "source" of the local turbulence  that can transport angular momentum outward. 
The computational focus has  employed  shearing boxes where a small radial  slab of the disc is considered (Balbus \& Hawley 1998, Balbus 2003). 
The focus on local boxes is also exacerbated by computational limitations.  It has not  yet been possible to simulate a global thin disc. 

Understanding the local physics is important but it is also noteworthy that 
some  paradigms of accretion discs 
involved buoyant magnetic loops 
  (Lynden-Bell 1969; FIeld \& Rogers 1993; Tout \& Pringle 1992; Johansen and Levin 2008)
  or outflows  (Blandford \& Payne 1982; 
K\"onigl 1989, even Colgate et al. 2001) involving large global   scale magnetic fields that transport the angular momentum.
 Maybe these are the primary modes of transport.
 
In short, although numerical simulations have become  fundamental tools
for progress, the limitations of tools can also limit the focus. Large scale and nonlocal features and outflows  may dominate the local transport.
There has been some renewed study along these  lines (Kuncic \& Bicknell 2007; Dobbie et al. 2008).  Hints that the role of large scale structures may
be quite fundamental is  discussed next.

\subsection{Role of  Large Scale Fields: Clues Even from quasi-local MRI Simulations}


Global simulations of thin accretion discs are a widely desired
community goal. Based on the sun and the observation of jets from accretion engines
and the existence of energetically significant coronae, 
large scale magnetic structures  and outflows must 
be playing some role in non-local angular momentum transport in non-self gravitating discs.
However even the study of quasi-local simulations is providing clues that non-local processes are important.
 
First, note that recently a systematic study of magnetic Prandtl number 
dependence of MRI saturation has begun (e.g. Fromang  et al. 2007; Davis et al. 2009).  
 This is an ongoing enterprise but  noteworthy is that stratified simulations seem to show less dependence on its value than unstratified solutions (Davis et al. 2009).
The stratified simulations also show less dependence on box size than do
unstratified simulations  (Pessah et al. 2007; Bodo et al. 2008).
(How the role of parasitic modes as a possible mechanism of MRI saturation (Pessah \& Goodman 2009) depends on box size for stratified vs. unstratified cases is uncertain.)

Since real systems are stratified, the weaker dependence of the saturated state on box properties is  step forward, but with stratification 
comes magnetic buoyancy and coronae. Large scale structures have the longest turbulent diffusion times and are most likely to survive the buoyant rise (Blackman \& Pessah 2009). Thus if stratification is important
then understanding what radial and vertical scales are large enough to fully capture
the dominant magnetic stress remains a question that more global simulations may be needed to answer.  Presently, the radial width of the Cartesian box simulations
are significantly smaller than the disc radius but magnetic coupling of different regions in the disc via magnetic loops can non-locally transport angular momentum (e.g. Lynden-Bell 1969; Blandford \& Payne 1982; Tout \& Pringle 1992; Field and Rogers 1993).

Another point is  that shearing box simulations show the generation of large scale
mean toroidal magnetic fields that last many orbit times and exhibit sign reversals
(Brandenburg et al. 1995, Lesur \& Ogilvie 2008, Davis et al. 2009).
These are seen in stratified simulations with and without periodic vertical boundaries,
and in unstratified simulations with periodic vertical boundaries (Lesur \& Ogilvie 2008).  The patterns indicate a large scale dynamo operating contemporaneously with the small scale dynamo.
When stratification is present,  there are global (or hemispheric)  pseudoscalars which depend on the product of rotation and density stratification that can supply  a mean magnetic field aligned EMF (e.g. Brandenburg \& Donner 1997; Brandenburg 1998; Tan \& Blackman 2004) .  But even in the unstratified cases, there seems to be
a mean field aligned EMF in sub-global but non-local regions of the disc.
  Any presence of a mean field aligned EMF is then a sub-global but non-local source of magnetic helicity or magnetic helicity flux of opposite sign between sectors. 
 All of this links  back to the LSD discussion of section 3.

Finally, note that shearing box simulations 
impose a steady-state by artificially enforcing the rotation profile with no actual 
accretion. This rotation is not subject to the back-reaction 
of the amplified field or flow dynamics.
Hubbard and Blackman (2009), argue that this may be too restrictive: 
In a real disc, energy in differential rotation is sustained  only by accretion 
itself. If a steady-state is to be maintained via turbulent transport alone
then there must be 100\%   power throughput from differential rotation to the turbulent cascade.
This is not guaranteed if large scale fields drain power.  A realistic 
steady-state solution would also have to  then incorporate stresses from large-scale fields.

\subsection{Coupling of Disc, Corona, and Jet}

That observations  reveal jets and  non-thermal emission best explained as emanating from  coronae in accretion engines most directly highlights the importance
 of large scale fields therein. There is debate as to whether or not the very global scale fields of jets are  accreted or produced in situ (Lovelace et al. 2009), but as emphasized, the sun, itself a stratified rotator,  produces its global scale fields in situ because the sign of these fields reverse.  Coronal loops are large scale with respect to the interior eddies but small
scale with respect to the global scale fields of coronal holes.
The formation of the global scale fields of coronal holes along which the solar wind propagates
occurs as some of the closed loops  open up (cite{sz}).
%
 Jets from accretion engines in 
young stellar objects, active galactic nuclei and 
gamma-ray bursts
or magnetic towers (\cite{pudritz04,lb03,um06})
may be the direct analogue to coronal holes.
These  large scale coronal fields
can be produced by the MDHD relaxation of smaller scale
loops emerging from a FDHD inside the rotator below (Blackman 2007)
If the helical field injected from below is  produced by a FDHD inside the rotator, the full description of the origin of the large scale fields involves a coupling between the FDHD and MDHD through the coronal boundary, which in turn is coupled to the MRI
in the accretion disc. 


Starting with the assumption that coronal luminosity from a turbulent
accretion disc results from buoyant magnetic structures that survive
turbulent shredding for at least one vertical density scale height, Blackman \& Pessah (2009)
derived lower limits on: ({\it i}) the scale of such magnetic
structures and ({\it ii}) the fraction of magnetic energy that needs
to be produced above this scale $l_{\rm c}$  within the disc to account for
observed values of coronal to bolometric luminosity. 
They considered  buoyant structures in pressure
equilibrium with the ambient medium but with an additional magnetic
energy contribution from scales above  $l_{\rm c}$, and thus a lower-than-ambient density.
The results  imply  that typical  coronal (non-thermal) to bolometric luminosity
ratios observed in AGN require the critical scale for buoyancy to robustly
exceed the  scale of the largest turbulent motions $l_t$ and that
double digit percentages of magnetic energy produced in the disc  reside 
in large scale structures (e.g. flux tubes) whose smallest scale exceeds $l_t$.

The above results are consistent with recent work highlighting
the importance of in situ large scale dynamos in feeding coronae 
(Blackman \& Field 2000; Vishniac 2009) and 
 further motivate larger domains in stratified MRI simulations
and determination of the magnetic energy and magnetic helicity spectra produced therein.

\section{Combining MFD and Accretion Theory into  a Single Mean Field Theory}

 Understanding the competition between diffusion of global  scale
magnetic fields in accretion discs, advection of flux, and amplification of flux is all part of the same question: how do mean magnetic fields evolve in accretion discs when the dynamically correct mean field transport coefficients are obtained for both the coupled  LSD and accretion dynamics? Ultimately, mean field accretion disc theory should be coupled to a MFD theory in a real disc as they are artificially separated 
components of what should be a single mean field theory.   A coupled theory will involve
separate the time evolution equations for mean velocity, surface density, 
mean magnetic field, and an energy equation.  

Some aspects of the coupled evolution of mean velocity field and
mean magnetic field in dynamo theory have been studied, though not part of  accretion
 theory (Blackman 1997;  Blackman et al.  2006; Courvoiser et al. 2009).
Ingredients of  a more formal dynamical mean field accretion disc theory are also emerging (Pessah et al. 2007; Hubbard \& Blackman 2009) which need
to be combined with lessons learned form MFD theory.
Campbell \& Caunt (1999) combined 20th century MFD theory with the SS73
accretion theory  closure to include local and large scale stresses. 
The SS73 type viscosity (via MRI) has also been used  as a closure for the turbulence from which  semi analytic MFDT is then derived to produce large scale fields in discs  to power outflows (e.g. Tan \& Blackman 2004). 20th century MFD theory has also been applied to MRI stratified simulations to explain observed cycle periods (Brandenburg et al. 1995;  Brandenburg \& Donner 1997; also Lesur \& Ogilvie 2008; Davis et al. 2009). 

Although none of the above approaches yet combine insight gained from understanding
nonlinear MFD theory via magnetic helicity conservation as discussed in section 3 with  accretion transport coefficients beyond that of SS73, there are indirect clues that the role of the $\alpha_{mag}$ term discussed in section 3.1 might be the driver of large scale field growth  in sheared stratified rotators rather than the $\alpha_{kin}$ term.  
This is based on (i) the sign of the $\alpha_{dyn}$ needed to model the parity and cycle periods seen in MRI simulations with a LSD theory 
 (Brandenburg \& Donner 1997; Brandenburg 1998, R\"udiger \& Pipin 2000;  Rekowski et al. 2000; Gressel 2009), (ii) the mechanism of of a magnetic  instability driven dynamo within stratified rotators (Spruit 2002), and (iii) the fact that the magnetic fluctuations seem to dominate the velocity fluctuations in all MRI simulations.   With solutions to LSD models
that match the accretion disk simulations,  the extent to which large scale fields
 transport angular momentum can be studied.

The coupling of the disc to the corona in a real system highlights the role of magnetic energy flux and magnetic helicity fluxes so  the dynamics of angular momentum transport in real discs are unlikely to be captured with unstratified simulations.

\section{Conclusions}

For any turbulent MHD  system in which there is both turbulence on short time scales but
ordered large scale patterns,  there has to  be a mean field theory that dynamically 
 captures essential physics and key principles in a useful framework.
 Note that MHD itself is already a mean field theory which averages out the noise of individual particle motions.  The question for turbulent rotators 
 involving  accretion and large scale dynamos is not
"is mean field theory is correct?"  but rather "do we have the correct mean field theory?"
Progress in large scale dynamo theory is farther along that accretion theory in this regard.

21st century mean field dynamo theory has incurred a paradigm shift compared to
that of the 20th century  by incorporating the dynamical evolution of magnetic helicity. This has led to  mean field models that successfully predict the non-linear saturation of the large
scale field growth seen in numerical simulations.  In that way, 21st century dynamo theory
has overcome an long standing challenge of 20th century dynamo theory. 
More work is still needed in applying the principles learned to realistic systems.

The analogous paradigm shift has not yet occurred in accretion theory.
 The single $\alpha_{ss}$ viscosity parameter of SS73 does not capture the quantitative
 properties of angular momentum transport in numerical simulations or 
 provide testable dynamical predictions   for either  local turbulent stresses produced by the MRI, or large scale magnetic stresses.  Some principles behind
a generalized semi-analytic accretion model must include are emerging, and
more effective closures than the $\alpha_{ss}$ prescription are showing some promise.

Ultimately, the need to incorporate both
the large scale magnetic field growth and its stresses along with those provided
by small scale field, implies that semi-analytic mean field dynamo theory and semi-analytic accretion disc theory are two components of what is
a unified mean field theory of accretion. Developing this unified theory is a challenging
and meaningful goal for the present century.

\acknowledgements
Thanks to K. Subramanian for a careful reading and comments and to O. Gressel for comments.
Thanks to the organizers  for a stimulating celebration and cheers to Axel for continued  fulfillment. 
I acknowledge NSF grants  AST-0406799, AST-0406823, and NASA grant ATP04 -0000-0016 and the LLE at UR. 


\begin{thebibliography}{}


\bibitem[Balbus 
\& Hawley(1998)]{1998RvMP...70....1B} Balbus, S.~A., \& Hawley, J.~F.\ 1998, Reviews of Modern Physics, 70, 1 


\bibitem[Balbus(2003)]{2003ARA&A..41..555B} Balbus, S.~A.\ 2003, \araa, 41, 555 


\bibitem[Balbus et al.(1994)]{bgh94} Balbus, S.~A., Gammie, 
C.~F., \& Hawley, J.~F.\ 1994, \mnras, 271, 197 

\bibitem[Beck et 
al. 1996]{beck96} Beck, R., Brandenburg, A., Moss, D., Shukurov, A., \& Sokoloff, D.\ 1996, \araa, 34, 155 

\bibitem[Bellan 2000]{bellan}  Bellan P.M., 2000, {\sl Spheromaks}, 
(Imperial College Press, London)

\bibitem[Berger 
\& Field(1984)]{1984JFM...147..133B} Berger, M.~A., \& Field, G.~B.\ 1984, Journal of Fluid Mechanics, 147, 133 



\bibitem[Blackman(1998a)]{1998ApJ...496L..17B} Blackman, E.~G.\ 1998a, \apjl, 
496, L17 

\bibitem[Blackman(1998b)]{1998MNRAS.299L..48B} Blackman, E.~G.\ 1998b, 
\mnras, 299, L48 




\bibitem[Blackman 
\& Field(2000)]{2000MNRAS.318..724B} Blackman, E.~G., \& Field, G.~B.\ 2000, \mnras, 318, 724 

\bibitem[Blackman 
\& Field(2002)]{2002PhRvL..89z5007B} Blackman, E.~G., \& Field, G.~B.\ 2002, Physical Review Letters, 89, 265007 


\bibitem[Blackman 
\& Brandenburg(2003)]{2003ApJ...584L..99B} Blackman, E.~G., \& Brandenburg, A.\ 2003, \apjl, 584, L99 

\bibitem[Blackman(2003)]{2003MNRAS.344..707B} Blackman, E.~G.\ 2003, 
\mnras, 344, 707 




\bibitem[Blackman(2004)]{B04} Blackman, E.~G.\ 2004, Plasma 
Physics and Controlled Fusion, 46, 423 

\bibitem[Blackman 
\& Field(2004)]{2004PhPl...11.3264B} Blackman, E.~G., \& Field, G.~B.\ 2004, Physics of Plasmas, 11, 3264 

\bibitem[Blackman et al.(2006)]{2006NewA...11..452B} Blackman, E.~G., 
Nordhaus, J.~T., \& Thomas, J.~H.\ 2006, New Astronomy, 11, 452 


\bibitem[Blackman(2007)]{2007NJPh....9..309B} Blackman, E.~G.\ 2007, New 
Journal of Physics, 9, 309 


\bibitem[Blackman et al.(2008)]{2008NewA...13..244B} Blackman, E.~G., 
Penna, R.~F., \& Varni{\`e}re, P.\ 2008, New Astronomy, 13, 244 




\bibitem[Blackman \& Pessah(2009)]{2009arXiv0907.2068B} Blackman, E.~G., \& Pessah, M.~E.\ 2009, arXiv:0907.2068,  ApJ, 704 L113





\bibitem[Blandford 
\& Payne(1982)]{1982MNRAS.199..883B} Blandford, R.~D., \& Payne, D.~G.\ 1982, \mnras, 199, 883 



\bibitem[Bodo et 
al.(2008)]{bodo2008} Bodo, G., Mignone, A., Cattaneo, F., Rossi, P., \& Ferrari, A.\ 2008, \aap, 487, 1 

\bibitem[Brandenburg et al.(1995)]{1995ApJ...446..741B} Brandenburg, A., 
Nordlund, A., Stein, R.~F., \& Torkelsson, U.\ 1995, \apj, 446, 741 

\bibitem[Brandenburg 
\& Donner(1997)]{1997MNRAS.288L..29B} Brandenburg, A., \& Donner, K.~J.\ 1997, \mnras, 288, L29 


\bibitem[Brandenburg(1998)]{1998tbha.conf...61B} Brandenburg, A.\ 1998, 
in "Theory of Black Hole Accretion Disks," edited by M.A. Abramowicz, G.Bjornsson, and J. E. Pringle. Cambridge University Press, 1998., p.61



\bibitem[Brandenburg(2001)]{2001ApJ...550..824B} Brandenburg, A.\ 2001, 
\apj, 550, 824 


\bibitem[Brandenburg \& Sandin (2004)]{bs04} Brandenburg, A., \& Sandin, C.\ 2004, \aap, 427, 13 



\bibitem[Brandenburg 
\& Subramanian(2005)]{2005PhR...417....1B} Brandenburg, A., \& Subramanian, K.\ 2005, Phys. Rep., 417, 1 


\bibitem[Brandenburg et al.(2008)]{2008ApJ...676..740B} Brandenburg, A., 
R{\"a}dler, K.-H., Rheinhardt, M., {K\"a}pyl{\"a}, P.~J.\ 2008, \apj, 676, 740 


\bibitem[Campbell 
\& Caunt(1999)]{1999MNRAS.306..122C} Campbell, C.~G., \& Caunt, S.~E.\ 1999, \mnras, 306, 122 

\bibitem[Cattaneo(1994)]{1994ApJ...434..200C} Cattaneo, F.\ 1994, \apj, 
434, 200 



\bibitem[Colgate et al.(2001)]{2001PhPl....8.2425C} Colgate, S.~A., Li, H., 
\& Pariev, V.\ 2001, Physics of Plasmas, 8, 2425




\bibitem[Courvoisier et al.(2009)]{2009arXiv0909.0721C} Courvoisier, A., 
Hughes, D.~W., \& Proctor, M.~R.~E.\ 2009, arXiv:0909.0721 


\bibitem[Davis et al.(2009)]{2009arXiv0909.1570D} Davis, S.~W., Stone, 
J.~M., \& Pessah, M.~E.\ 2009, arXiv:0909.1570 

\bibitem[Dobbie et al.(2009)]{2009arXiv0903.3274D} Dobbie, P.~B., Kuncic, 
Z., Bicknell, G.~V., \& Salmeron, R.\ 2009, arXiv:0903.3274 



\bibitem[Field 
\& Rogers(1993)]{1993ApJ...403...94F} Field, G.~B., \& Rogers, R.~D.\ 1993, \apj, 403, 94 

\bibitem[Frank et al.(2002)]{2002apa..book.....F} Frank, J., King, A., 
\& Raine, D.~J.\ 2002, Accretion Power in Astrophysics, by Juhan Frank and Andrew King and Derek Raine, pp.~398.~ISBN 0521620538.~Cambridge, UK: Cambridge University Press, February 2002.,  



\bibitem[Fromang et 
al.(2007)]{2007AA...476.1123F} Fromang, S., Papaloizou, J., Lesur, G., \& Heinemann, T.\ 2007, \aap, 476, 1123 


\bibitem[Gressel 2009]{2007AA...476.1123F} Gressel, O, 2009, to be submitted to MNRAS 

\bibitem[Hubbard 
\& Blackman(2008)]{2008MNRAS.390..331H} Hubbard, A., \& Blackman, E.~G.\ 2008, \mnras, 390, 331 


\bibitem[Hubbard 
\& Blackman(2009)]{2009MNRAS.398..931H} Hubbard, A., \& Blackman, E.~G.\ 2009, \mnras, 398, 931 



\bibitem[Ji \& Prager 2002]{jiprager} 
 Ji H., \&  Prager S.C., 2002,  {Magnetohydrodynamics} {\bf 38}, 191 
 
\bibitem[Johansen 
\& Levin(2008)]{2008A&A...490..501J} Johansen, A., \& Levin, Y.\ 2008, \aap, 490, 501 


\bibitem[K{\"a}pyl{\"a} et 
al.(2008)]{2008AA...491..353K} K{\"a}pyl{\"a}, P.~J., Korpi, M.~J., \& Brandenburg, A.\ 2008, \aap, 491, 353 

\bibitem[] {}Kleeorin, N. I. \& Ruzmaikin, A. A., 1982 
 Magnetohydrodynamics 18, 116

\bibitem[Krause \& R\"adler 1980]{kr80}Krause F. and RŠdler K. H. Mean Field Magnetohydrodynamics and Dynamo Theory 1980, Pergamon Press.

\bibitem[Konigl(1989)]{1989ApJ...342..208K} Konigl, A.\ 1989, \apj, 342, 
208 



\bibitem[Kulsrud(2005)]{2005LNP...664...69K} Kulsrud, R.~M.\ 2005, in {\it Cosmic 
Magnetic Fields},  Eds: R. Wielebinski, R. Beck, Springer Lecture Notes in Physics, vol. 664, p.69


 \bibitem[Kuncic 
\& Bicknell(2007)]{2007Ap&SS.311..127K} Kuncic, Z., \& Bicknell, G.~V.\ 2007, \apss, 311, 127 


\bibitem[Lesur \& Ogilvie(2008)]{2008A...488..451L} Lesur, G., \& Ogilvie, G.~I.\ 2008, A\&A, 488, 451 

\bibitem[Lovelace et al.(2009)]{2009ApJ...701..885L} Lovelace, R.~V.~E., 
Rothstein, D.~M., \& Bisnovatyi-Kogan, G.~S.\ 2009, \apj, 701, 885 

\bibitem[Lynden-Bell(1969)]{1969Natur.223..690L} Lynden-Bell, D.\ 1969, 
\nat, 223, 690 

\bibitem[Lynden-Bell (2003)]{lb03}  Lynden-Bell, D.,  2003,
MNRAS, { 341} 1360 






\bibitem[Moffatt (1978)]{moffatt} Moffatt, H.~K.\ 1978, 
Cambridge, England, Cambridge University Press, 1978.~353 p.,  


\bibitem[Ogilvie(2003)]{2003MNRAS.340..969O} Ogilvie, G.~I.\ 2003, \mnras, 
340, 969 



\bibitem[Parker 1979 ]{parker} Parker, E.~N.\ 1979, Oxford, 
Clarendon Press; New York, Oxford University Press, 1979, 858p

\bibitem[Pessah et al.(2006)]{2006PhRvL..97v1103P} Pessah, M.~E., Chan, 
C.-K., \& Psaltis, D.\ 2006, Physical Review Letters, 97, 221103 

\bibitem[Pessah et al.(2007)]{2007ApJ...668L..51P} Pessah, M.~E., Chan, 
C.-k., \& Psaltis, D.\ 2007, \apjl, 668, L51 



\bibitem[Pessah et al.(2008)]{2008MNRAS.383..683P} Pessah, M.~E., Chan, 
C.-K., \& Psaltis, D.\ 2008, \mnras, 383, 683 


\bibitem[Pessah 
\& Goodman(2009)]{2009ApJ...698L..72P} Pessah, M.~E., \& Goodman, J.\ 2009, \apjl, 698, L72 

\bibitem[]{pfl} Pouquet A.,   Frisch U.,   L\'eorat J., 1976,   J. Fluid Mech., { 77}  321 


\bibitem[Pudritz (2004)]{pudritz04} Pudritz, R.~E.\ 2004, \apss, 292, 471 






\bibitem[Rogachevskii 
\& Kleeorin(2003)]{2003PhRvE..68c6301R} Rogachevskii, I., \& Kleeorin, N.\ 2003, PRE, 68, 036301 




\bibitem[R{\"u}diger 
\& Pipin(2000)]{2000A&A...362..756R} R{\"u}diger, G., \& Pipin, V.~V.\ 2000, \aap, 362, 756 


\bibitem[Schekochihin et al.(2008)]{2008arXiv0810.2225S} Schekochihin, 
A.~A., et al.,
J.~C., Rogachevskii, I., \& Yousef, T.~A.\ 
2008, arXiv:0810.2225 


\bibitem[ Schrijver  \&  Zwaan 2000]{sz}
 Schrijver  C.J. \&  Zwaan C., 2000, {\it Solar and Stellar Magnetic Activity},
(Cambridge: Cambridge Univ. Press)

\bibitem[Shakura 
\& Syunyaev(1973)]{ss73} Shakura, N.~I., \& Syunyaev, R.~A.\ 1973, \aap, 24, 337 (SS73)

\bibitem[Shi et al.(2009)]{2009arXiv0909.2003S} Shi, J.-M., Krolik, J.~H., 
\& Hirose, S.\ 2009, arXiv:0909.2003, submitted to ApJ. 


\bibitem[Shukurov 2002]{shukurov02} Shukurov, A.\ 2002, \apss, 281, 285 



\bibitem[Snellman et al.(2009)]{2009arXiv0906.1200S} Snellman, J.~E., 
K{\"a}pyl{\"a}, P.~J., Korpi, M.~J., 
\& Liljestr{\"o}m, A.~J.\ 2009, arXiv:0906.1200 

\bibitem[Spruit(2002)]{2002A&A...381..923S} Spruit, H.~C.\ 2002, \aap, 381, 923 


\bibitem[Sridhar 
\& Subramanian(2009)]{2009PhRvE..79d5305S} Sridhar, S., \& Subramanian, K.\ 2009a, Phys. Rev. E., 79, 045305 

\bibitem[Sridhar 
\& Subramanian(2009)]{2009arXiv0906.3073S} Sridhar, S., \& Subramanian, K.\ 2009b, arXiv:0906.3073 


\bibitem[Subramanian 
\& Brandenburg(2006)]{2006ApJ...648L..71S} Subramanian, K., \& Brandenburg, A.\ 2006, \apjl, 648, L71 



\bibitem[Sur et al.(2007)]{2007MNRAS.377..874S} Sur, S., Shukurov, A., 
\& Subramanian, K.\ 2007, \mnras, 377, 874 

\bibitem[Sur 
\& Subramanian(2009)]{2009MNRAS.392L...6S} Sur, S., \& Subramanian, K.\ 2009, MNRAS, 392, L6 

\bibitem[Tan 
\& Blackman(2004)]{2004ApJ...603..401T} Tan, J.~C., \& Blackman, E.~G.\ 2004, \apj, 603, 401 

\bibitem[Taylor(1986)]{1986RvMP...58..741T} Taylor, J.~B.\ 1986, Reviews of 
Modern Physics, 58, 741 

\bibitem[Tout 
\& Pringle(1992)]{1992MNRAS.259..604T} Tout, C.~A., \& Pringle, J.~E.\ 1992, \mnras, 259, 604 


\bibitem[Treves et al.(1989)]{1989AdSAC...5.....T} Treves, A., Maraschi, 
L., 
\& Abramowicz, M.\ 1989, {\it Accretion: A Collection of  Influential Papers}
Advanced Series in Astrophysics and Cosmology, 5,
(World Scientific: Singapore)



\bibitem
[Uzdensky \& MacFadyen (2006)]
{um06} Uzdensky D.A., 
\& MacFadyen,  A.I., 2006, ApJ, {\bf 647} 1192


\bibitem[Vall{\'e}e 2004]{vallee04} Vall{\'e}e, J.~P.\ 2004, 
New Astronomy Review, 48, 763 

\bibitem[Vishniac 
\& Brandenburg(1997)]{1997ApJ...475..263V} Vishniac, E.~T., \& Brandenburg, A.\ 1997, \apj, 475, 263 

\bibitem[Vishniac 
\& Cho(2001)]{2001ApJ...550..752V} Vishniac, E.~T., \& Cho, J.\ 2001, \apj, 550, 752 

\bibitem[Vishniac(2009)]{2009ApJ...696.1021V} Vishniac, E.~T.\ 2009, \apj, 
696, 1021 

\bibitem[Rekowski et 
al.(2000)]{2000A&A...353..813R}  Rekowski, M.~v., R{\"u}diger, G., \& Elstner, D.\ 2000, \aap, 353, 813 



\bibitem[Wang 
\& Sheeley (2003)]{ws03} Wang, Y.-M., \& Sheeley, N.~R., Jr.\ 2003, \apj, 599, 1404 


\bibitem[Yousef et al.(2008)]{2008PhRvL.100r4501Y} Yousef, T.~A.
et al., 
 2008, Phys Rev. Lett., 100, 184501 


 
 

\end{thebibliography}
\end{document}